\documentclass[a4paper,USenglish,cleveref, autoref]{lipics-v2019}

\usepackage{bm} 

\input{macros}

\title{Normalization by Evaluation for Call-by-Push-Value and Polarized Lambda-Calculus}

\titlerunning{NbE for CBPV and Focusing}

\author{Andreas Abel}
  {Department of Computer Science and Engineering,
    Chalmers and Gothenburg University,
    Sweden \and
    \url{www.cse.chalmers.se/~abela}
  }
  {andreas.abel@gu.se}
  {https://orcid.org/0000-0003-0420-4492}
  {VR Grant 2014-04864 \emph{Termination
      Certificates for Dependently-Typed Programs and Proofs via
      Refinement Types}
  }

\author{Christian Sattler}
  {Department of Computer Science and Engineering,
    Chalmers and Gothenburg University,
    Sweden
  }
  {}
  {}
  {}

\authorrunning{A. Abel 
and C. Sattler}

\Copyright{Andreas Abel 
 and Christian Sattler}



\category{}

\relatedversion{}

\supplement{\url{https://andreasabel.github.io/ipl/html/NfModelMonad.html}}

\acknowledgements{The first author took inspiration from some
  unpublished notes by Thorsten Altenkirch titled
\emph{Another topological completeness proof for intuitionistic logic}
received by email on 16th March 2000.
Thorsten in turn credits his inspiration to an ALF proof by Thierry
Coquand.}

\nolinenumbers 

\hideLIPIcs  

\EventEditors{Jakob Rehof and XXX}
\EventNoEds{2}
\EventLongTitle{Formal Structures for Computation and Deduction}
\EventShortTitle{FSCD 2019}
\EventAcronym{FSCD}
\EventYear{2019}
\EventDate{June 24--30, 2019}
\EventLocation{Dortmund, Germany}
\EventLogo{}
\SeriesVolume{}
\ArticleNo{}

\begin{document}

\maketitle

\begin{abstract}
We observe that normalization by evaluation for simply-typed
lambda-calculus with weak coproducts can be carried out in a weak
bi-cartesian closed category of presheaves equipped with a monad that
allows us to perform case distinction on neutral terms of sum type.
The placement of the monad influences the normal forms we obtain: for
instance, placing the monad on coproducts gives us eta-long beta-pi
normal forms where pi refers to permutation of case distinctions out
of elimination positions.  We further observe that placing the monad
on every coproduct is rather wasteful, and an optimal placement of the
monad can be determined by considering polarized simple types inspired
by focalization.  Polarization classifies types into positive and
negative, and it is sufficient to place the monad at the embedding of
positive types into negative ones.  We consider two calculi based on
polarized types: pure call-by-push-value (CBPV) and polarized
lambda-calculus, the natural deduction calculus corresponding to
focalized sequent calculus.  For these two calculi, we present
algorithms for normalization by evaluation.  We further discuss
different implementations of the monad and their relation to existing
normalization proofs for lambda-calculus with sums.  Our developments
have been partially formalized in the Agda proof assistant.

\end{abstract}

\section{Introduction}
\label{sec:intro}

The idea behind \emph{normalization by evaluation} (NbE) is to utilize a
standard interpreter, usually evaluating closed terms, to compute the
normal form of an open term.
The normal form is obtained by a type-directed \emph{reification}
procedure after evaluating the open term to a semantic value, mapping
(\emph{reflecting})
the free variables to corresponding \emph{unknowns} in the semantics.
The literal use of a standard interpreter
can be achieved for the pure simply-typed lambda-calculus
\cite{bergerSchwichtenberg:lics91,filinski:semaccounttdpe}
by modelling uninterpreted base types as sets of neutral (aka atomic)
terms, or more precisely, as presheaves or sets of neutral term
families, in order to facilitate fresh bound variable generation during
reification of functions to lambdas.
Thanks to $\eta$-equality at function types, free variables of
function type can be reflected into the semantics as functions
applying the variable to their reified argument,
forming a neutral term.  This mechanism provides us with unknowns of
function type which can be faithfully reified to normal forms.

For the extension to sum types (logically, disjunctions, and categorically,
weak coproducts), this reflection trick does not work anymore.  A
semantic value of binary sum type is either a left or a right
injection, but the decision between left or right cannot be taken at
reflection time, since a variable of sum type does not provide us with
such information.  A literal standard interpreter for closed terms can
thus no longer be used for NbE;  instead, we can utilize a monadic
interpreter.  When the interpreter attempts a case distinction on an
unknown of sum type, it asks an oracle whether the unknown is a left
or a right injection.  The oracle returns one of these alternatives,
wrapping a new unknown in the respective injection.  The communication with the
oracle can be modeled in a \emph{monad} $\C$,
which records the questions asked and the continuation of the
interpreter for each of the possible answers.
A monadic semantic value is thus a \emph{case tree} where the leaves
are occupied by non-monadic values \cite{altenkirchUustalu:flops04}.
In this article we only consider weak sum types, producing non-unique
normal forms, where it does not
matter in which order the questions are asked (commuting case splits),
and whether the same
question is asked several times (redundant case splits).
The model would need refinement for strong, extensional sums
\cite{altenkirchDybjerHofmannScott:lics01,altenkirchUustalu:flops04,balatDiCosmoFiore:popl04,barral:PhD,scherer:popl17}.

Filinski \cite{filinski:tlca01} studied NbE for Moggi's computational lambda
calculus \cite{moggi:infcomp91}, shedding light on the difference
between call-by-name (CBN) and call-by-value (CBV) NbE, where Danvy's
type-directed partial evaluation \cite{danvy:popl96} falls into the
latter class.  The contribution of the computational lambda calculus
is to make explicit where the monad is invoked during monadic
evaluation, and this placement of the monad carries over to the NbE
setting.  Moggi's studies were continued by Levy \cite{levy:hosc06}
who designed the call-by-push-value (CBPV) lambda-calculus to embed
both the CBN and CBV lambda calculus.

In this work, we formulate NbE for CBPV
(Section~\ref{sec:cbpv}), with the aim to investigate
later whether CBN and CBV NbE can be recovered from CBPV NbE via the
standard translations of the CBN and CBV calculi into CBPV.

In contrast to the normal forms of CBN NbE, which is the algorithmic
counterpart of the completeness proof for intuitionistic propositional
logic (IPL) using Beth models, CBPV NbE gives us more restrained
normal forms, where the production of a value via injections cannot be
interrupted by more questions to the oracle.  In the research field of
focalization \cite{andreoli:focusing,liangMiller:csl07} we speak of
\emph{chaining non-invertible introductions}.
Invertible introductions are already chained in NbE thanks to
extensionality ($\eta$) for function, and more generally, negative
types.  Non-invertible eliminations are also happening in a chain when
building neutrals.  What is missing from the picture is the chaining
of invertible eliminations, i.e., case distinctions and, more
generally, pattern matching.  The picture is completed by extending
NbE to \emph{polarized lambda calculus}
\cite{zeilberger:PhD,brockNannestadSchuermann:lpar10,espiritoSanto:entcs17}
in Section~\ref{sec:fipl}.

In our presentation of the various lambda calculi we ignore the
concrete syntax, only consider the abstract syntax obtained by the
Curry-Howard-Isomorphism.  A term is simply a derivation tree whose
nodes are rule invocations.  Thus, a intrinsically typed,
nameless syntax is most natural,
and our syntactic classes are all presheaves over the category of
typing contexts and renamings.  The use of presheaves then smoothly
extents to the semantic constructions
\cite{catarina:csl93,altenkirchHofmannStreicher:ctcs95}.

Concerning the presentation of polarized lambda calculus, we depart from
Zeilberger \cite{zeilberger:PhD} who employs \emph{a priori}
infinitary syntax, modelling a case tree as a meta-level function mapping
well-typed patterns to branches.  Instead, we use a graded monad
representing complete pattern matching over a newly added hypothesis,
which is in spirit akin to Filinski's \cite[Section~4]{filinski:tlca01}
and Krishnaswami's \cite{krishnaswami:popl09} treatment of eager
pattern matching using a separate context of variables to be matched
on.

Our design choices were guided by an Agda formalization of sections 2
(complete) and 4 (partial), available at \url{https://github.com/andreasabel/ipl}.
Agda was particularly helpful to correctly handle the renamings
abundantly present when working with presheaves.


















\section{Normalization by Evaluation for the Simply-Typed Lambda
  Calculus with Sums}
\label{sec:lambda}

In this section, we review the normalization by evaluation (NbE)
argument for the simply-typed lambda calculus (STLC) with weak sums, setting
the stage for the later sections.
We work in a constructive type-theoretic meta-language, with the basic judgement
$t : T$ meaning that object $t$ is an inhabitant of type $T$.
However, to avoid confusion with object-level types such as the simple
types of lambda calculus, we will refer to meta-level types as
\emph{sets}.  Consequently, the colon $:$
takes the role of elementhood $\in$ in set theory, and we are free to
reuse the symbol $\in$ for other purposes.

\subsection{Contexts and indices}

We adapt a categorical aka de Bruijn style for the abstract syntax of
terms, which we conceive as intrinsically well-typed.  In de Bruijn
style, a context $\Gamma$ is just a snoc list of simple types $A$,
meaning we write context extension as $\cext \Gamma A$, and the empty
context as $\cempty$.  Membership \fbox{$A \in \Gamma$} and sublist
relations \fbox{$\Gamma \sublist \Delta$} are given inductively by the
following rules:
\begin{gather*}
  \nru{\tzero}{}{A \in \cext \Gamma A}
\qquad
  \nru{\tsuc}{A \in \Gamma}{A \in \cext \Gamma B}
\qquad
  \nru{\tdone}{}{\cempty \sublist \cempty}
\qquad
  \nru{\tlift}
    {\Gamma \sublist \Delta}
    {\cext \Gamma A \sublist \cext \Delta A}
\qquad
  \nru{\tweak}
    {\Gamma \sublist \Delta}
    {\Gamma \sublist \cext \Delta A}
\end{gather*}
We consider the rules as introductions of the indexed types
$\_{\in}\_$ and $\_{\sublist}\_$ and the rule names as constructors.
For instance, $\suc \tzero : A \in \cext {\cext \Gamma A} B$ for any
$\Gamma$, $A$, and $B$; and if we read $\tsuc^n\,\tzero$ as unary
number $n$, then $x : A \in \Gamma$ is exactly the (de Bruijn) index
of $A$ in $\Gamma$.

We can define
\fbox{$\tid : \Gamma \sublist \Gamma$} and \fbox{$\_{\rcomp}\_ : \Gamma
\sublist \Delta \to \Delta \sublist \Phi \to \Gamma \sublist \Phi$}
by recursion, meaning that the (proof-relevant)
sublist relation is reflexive and transitive.
Thus, lists $\Gamma$ form a category
$\Cxt$ with morphisms $\tau : \Gamma \sublist \Delta$, and the category laws
hold propositionally, e.g., we have $\id \rcomp \tau \equiv \tau$ in
propositional equality for all morphisms $\tau$.
The singleton weakening $\twk^A_\Gamma : \Gamma \sublist \cext \Gamma A$,
also written $\twk^A$ or $\twk$, is defined by $\twk = \tweak\,\tid$.

The category $\Cxt$ allows us to consider $A{\in}\_$ as a presheaf over
$\Cxt\op$ for any $A$, witnessed by $\treindex : \Gamma \sublist
  \Delta \to A \in \Gamma \to A \in \Delta$, which is the morphism
part of functor $A{\in}\_$ from $\Cxt$ to $\Set$, mapping object $\Gamma$ to the
set $A \in \Gamma$ of the indices of $A$ in $\Gamma$.
The associated functor laws $\treindex\,\tid\,x \equiv x$
and $\treindex\,\tau_2\,(\treindex\,\tau_1\,x) \equiv
\treindex\,(\tau_1 \rcomp \tau_2)\,x$ hold propositionally.

\subsection{STLC and its normal forms}

Simple types shall be distinguished into positive types $P$ and
negative types $N$, depending on their root type former.  Function
($\To$) and product types ($\times$ and $\tyone$) are negative,
while base types ($\tybase$) and sum types ($+$ and $\tyzero$) are positive.
\[
\begin{array}{lrl@{\qquad}l}
  A,B,C & ::= & P \mid N & \mbox{simple types} \\
  P     & ::= & \tyzero \mid A + B \mid \tybase & \mbox{positive types} \\
  N     & ::= & \tyone \mid A \times B \mid A \To B
    & \mbox{negative types} \\
\end{array}
\]
Intrinsically well-typed lambda-terms, in abstract syntax,
are just inhabitants $t$ of the indexed set \fbox{$A \from \Gamma$},
inductively defined by the following rules.
\begin{gather*}
  \nru{\tvar}{A \in \Gamma}{A \from \Gamma}
\qquad
  \nru{\tabs}{B \from \cext \Gamma A}{A \To B \from \Gamma}
\qquad
  \nru{\tapp}
    {A \To B \from \Gamma \qquad A \from \Gamma}
    {B \from \Gamma}
\\[2ex]
  \nru{\tunit}{}{\tyone \from \Gamma}
\qquad
  \nru{\tpair}
    {A_1 \from \Gamma \qquad A_2 \from \Gamma}
    {A_1 \times A_2 \from \Gamma}
\qquad
  \nru{\tprj_i}
    {A_1 \times A_2 \from \Gamma}
    {A_i \from \Gamma}
\\[2ex]
  \nru{\tinj_i}
    {A_i \from \Gamma}
    {A_1 + A_2 \from \Gamma}
\qquad
  \nru{\tcase}
    {A_1 + A_2 \from \Gamma
      \qquad B \from \cext \Gamma A_1
      \qquad B \from \cext \Gamma A_2
    }{B \from \Gamma}
\qquad
  \nru{\tabort}
    {\tyzero \from \Gamma}
    {B \from \Gamma}
\end{gather*}
The skilled eye of the reader will immediately recognize the proof
rules of intuitionistic propositional logic (IPL) under the
Curry-Howard isomorphism, where $A \from \Gamma$ is to be read as
``$A$ follows from $\Gamma$''.  Using shorthand
\fbox{$\vv n = \tvar\,(\tsuc^n\,\tzero)$} for the $n$th variable,
a term such as
\(
  \tabs\,(\tabs\,(\tpair\,
    \vv1\,
    (\tabs\,(\tapp\,\vv1\,\vv0))))
\)
could in concrete syntax be rendered as
\(
  \lambda x.\,\lambda y.\,(x,\,\lambda z.\,y\,z)
  .
\)
We leave the exact connection to a printable
syntax of the STLC to the imagination of the reader,
as we shall not be concerned with considering
concrete terms in this article.

Terms of type $A$ form a presheaf
$A{\from}\_$ as witnessed by the standard weakening operation%
\footnote{Here, $\tren$ is short for \emph{renaming}, but in a nameless
calculus we should better speak of \emph{reindexing}, which could, a
bit clumsily, be also abbreviated to $\tren$.}
\fbox{$\tren : \Gamma \sublist \Delta \to A \from \Gamma \to A \from \Delta$}
defined by recursion over $t : A \from \Gamma$,
and functor laws for $\tren$ analogously to $\treindex$.

Normal forms%
\footnote{%
There is also a stronger
notion of normal form, requiring that two \emph{extensionally equal}
lambda-terms, \ie, those that denote the
same set-theoretical function, have the same normal form
\cite{mitchell:foundations,altenkirchDybjerHofmannScott:lics01,scherer:popl17}.
Such normal forms do not have a simple inductive definition, and we
shall not consider them in this article.
}
are logically characterized as those fulfilling the
\emph{subformula} property \cite{prawitz:natded,matthes:shortproofs}.
Normal forms \fbox{$n : \tNf A\, \Gamma$} are mutually defined with
\emph{neutral} normal forms \fbox{$u : \Ne A \Gamma$}.
In the following inductive definition, we reuse the rule names from the
term constructors.
\begin{gather*}
  \nru{\tvar}{A \in \Gamma}{\Ne A \Gamma}
\qquad
  \nru{\tabs}{\NfC B {\cext \Gamma A}}{\NfP {A \To B} \Gamma}
\qquad
  \nru{\tapp}
    {\NeP {A \To B} \Gamma \qquad \Nf A \Gamma}
    {\Ne B \Gamma}
\qquad
  \nru{\tne}{\Ne \tybase \Gamma}{\Nf \tybase \Gamma}
\\[2ex]
  \nru{\tunit}{}{\Nf \tyone \Gamma}
\qquad
  \nru{\tpair}
    {\Nf {A_1} \Gamma \qquad \Nf {A_2} \Gamma}
    {\NfP {A_1 \times A_2} \Gamma}
\qquad
  \nru{\tprj_i}
    {\NeP {A_1 \times A_2} \Gamma}
    {\Ne {A_i} \Gamma}
\\[2ex]
  \nru{\tinj_i}
    {\Nf {A_i} \Gamma}
    {\NfP {A_1 + A_2} \Gamma}
\quad
  \nru{\tcase}
    {\NeP {A_1 + A_2} \Gamma
      \quad \NfC P {\cext \Gamma A_1}
      \quad \NfC P {\cext \Gamma A_2}
    }{\Nf P \Gamma}
\quad
  \nru{\tabort}
    {\Ne \tyzero \Gamma}
    {\Nf P \Gamma}
\end{gather*}
These rules only allow the elimination of \emph{neutrals}; this
restriction guarantees
the subformula property and prevents any kind of computational
($\beta$) redex.
The new rule $\tne$ embeds $\tNe$ into $\tNf$,
but only at base types $\tybase$ \cite[Section~3.3]{altenkirchHofmannStreicher:ctcs95}.
Further, case distinction via $\tcase$ and $\tabort$
is restricted to positive types $P$.
As a consequence, our normal forms are
\emph{$\eta$-long}, meaning that any normal inhabitant of a negative
type is a respective introduction ($\tabs$, $\tunit$, or $\tpair$).
This justifies the attribute \emph{negative} for these types:
the construction of their inhabitants proceeds mechanically, without
any choices.  In contrast, constructing an inhabitant of a
\emph{positive} type involves choice: whether case distinction is
required, and which introduction to pick in the end ($\tinj_1$ or $\tinj_2$).

Needless to say, $\tNe\,A$ and
$\tNf\,A$ are presheaves, \ie, support reindexing with $\tren$ just as terms do.
From a normal form we can extract the term via an overloaded function
\fbox{$\tm\_ : \Nf A \Gamma \to A \from \Gamma$} and
\fbox{$\tm\_ : \Ne A \Gamma \to A \from \Gamma$}
that discards constructor $\tne$ but keeps all other constructors.
This erasure function naturally commutes with reindexing, making it a
natural transformation between the presheaves $\tNf\,A$ ($\tNe\,A$,
resp.) and $A{\from}\_$.  We shall simply write, for instance,
$\tNf\,A \todot A{\from}\_$ for such presheaf morphisms.
(The point on the arrow is mnemonic for \emph{pointwise}.)
Slightly abusive, we shall extend this notation to $n$-ary
morphisms, \eg, write $\A \todot \B \todot \C$ for
$\forall \Gamma.\ \A\,\Gamma \to (\B\,\Gamma \to \C\,\Gamma)$.

\begin{remark*}
While the coproduct eliminations $\tcase$ and $\tabort$ are limited to
normal forms of positive types $P$, their extension $\tcase^B$ and
$\tabort^B$ to negative types
is admissible, for instance:
\[
\begin{array}{l@{~}lcl}
  \multicolumn 4 l {\tabort^B ~:~ \tNe\, 0 \todot \tNf\,B} \\
  \tabort^{\tyone}     & u & = & \tunit \\
  \tabort^P           & u & = & \tabort\;u \\
\end{array}
\qquad
\begin{array}{l@{~}lcl}
  \\
  \tabort^{A \times B} & u & = & \tpair\,(\tabort^A\,u)\,(\tabort^B\,u) \\
  \tabort^{A \To B}    & u & = & \tabs\,(\tabort^B\,(\tren\,\twk^A\,u)) \\
\end{array}
\]
$\tcase$ generalizes analogously, with a bit of care when weakening
the branches.
\end{remark*}




\subsection{Normalization}

Normalization is concerned with finding a normal form
$n : \Nf A \Gamma$ for each term $t : A \from \Gamma$.
The normal form should be \emph{sound},
\ie, $\tm n \cong t$
with respect to a equational theory $\cong$ on terms
(see \cref{sec:eq}).
Further, normalization should decide $\cong$, \ie,
terms $t,t'$ with $t \cong t'$ should have the same normal form $n$.
In this article, we implement only the normalization function
\fbox{$\tnorm : A \from \Gamma \to \Nf A \Gamma$}
with proving its soundness and completeness.
From a logical perspective, we will compute for each derivation of $A
\from \Gamma$ a normal derivation $\Nf A \Gamma$.



%
\emph{Normalization by evaluation} (NbE)
\fbox{\(
  \tnorm\, (t : A \from \Gamma)
  = \down A \dent t {\idenv\Gamma}
\)}
decomposes normalization into
\emph{evaluation}
\nofbox{$\denty\_ : (t : A \from \Gamma) \to \den \Gamma \todot
  \den A$}
in the identity environment \nofbox{$\idenv\Gamma : \den \Gamma \Gamma$}
followed by
\emph{reification}
\nofbox{$\down A : \den A \todot \tNf A$}
(aka quoting).
The role of evaluation is to produce from a term the corresponding
semantic (\ie, meta-theoretic) function, which is finally reified to a normal form.
Since we are evaluating open terms $t$, we need to supply an environment
$\idenv\Gamma$ which will map the free indices of $t$ to corresponding
\emph{unknowns}.  To accommodate unknowns in the semantics, types $A$
are mapped to presheaves $\den A$ (rather than just sets), and in
particular each base type $\tybase$ is mapped to the presheaf
$\tNe\,\tybase$ with the intention that the neutrals take the role of
the unknowns.  The mapping $\up A : \tNe\,A \todot \den A$ from
neutrals to unknowns is called \emph{reflection} (aka unquoting), and defined mutually
with reification by induction on type $A$.

At this point, let us fix some notation for sets to prepare for some
constructions of presheaves.  Let $\bone$ denote the unit set and
$\ttempty$ its unique inhabitant, $\bzero$ the empty set and
$\tmagic : \bzero \to T$ the \emph{ex falsum quod libet} elimination
into any set $T$.  Given sets $S_1$ and $S_2$, their Cartesian product
is written $S_1 \btimes S_2$ with projections $\pi_i : S_1 \btimes S_2
\to S_i$, and their disjoint sum $S_1 \bplus S_2$ with injections
$\iota_i : S_i \to S_1 \bplus S_2$ and elimination $[f_1,f_2] : S_1 \bplus
S_2 \to T$ for arbitrary $f_i : S_i \to T$.

Presheaves (co)products $\hat0$, $\hat1$, $\hatplus$, and $\hattimes$ are
constructed pointwise, \eg, $\hat0\,\Gamma = \bzero$, and given two
presheaves $\A$ and $\B$,
$(\A \hatplus \B)\,\Gamma = \A\,\Gamma \bplus \B\,\Gamma$.  For the
exponential of presheaves, however, we need the \emph{Kripke function
  space}
$(\A \hatto \B)\,\Gamma = \forall \Delta.\, \Gamma \sublist \Delta
\to \A\,\Delta \to \B\,\Delta$.

We will interpret simple types $A$ as corresponding presheaves
$\den A$.  Let us start with the negative types, defining
reflection $\up A : \tNe\, A \todot \den A$
and reification $\down A : \den A \todot \tNf A$ along the way.
\[
\begin{array}{lcl}
  \den   \tyone             & = & \hatone \\
  \upG   \tyone \; u        & = & \ttempty \\
  \downG \tyone \; \ttempty & = & \tunit \\
\end{array}
\qquad\qquad\qquad
\begin{array}{lcl}
  \den   {A \times B}       & = & \den A \hattimes \den B \\
  \upG   {A \times B} \; u  & = & (\upG A (\prj1 u),\;\upG B (\prj2 u)) \\
  \downG {A \times B} (a,b) & = & \tpair\,(\downG A a)\,(\downG B b) \\
\end{array}
\]
\[
\begin{array}{lcl}
  \den {A \To B}    & = & \den A \hatto \den B \\
  \upG   {A \To B}    \; u \; (\tau : \Gamma \sublist \Delta) \; (a : \den A \Delta)
     & = & \up[\Delta] B (\tapp\;(\tren\,\tau\, u)\;(\down[\Delta] A a)) \\
  \downG {A \To B}    f        & = & \tabs\,(\down[\cext \Gamma A] B
    (f\;\twk^A_\Gamma\;\tfresh^A_\Gamma))
\end{array}
\]
In the reification at function types
$\down{A \To B}$,
the renaming $\twk^A_\Gamma : \Gamma \sublist \cext \Gamma A$
makes room
for a new variable of type $A$, which is reflected into $\den A$ by
\fbox{$\tfresh^A_\Gamma = \up[\cext \Gamma A] A {\vv0} : \den A (\cext \Gamma A)$}.
The ability to introduce fresh variables into a context, and to use
semantic objects such as $f : \den{A \To B}\Gamma$ in a such extended
context, is the reason for utilizing presheaves instead of just sets
as semantic types.

Note also that in the equation for $\up{A \To B}$,
the neutral $u : \Ne A \Gamma$ is
transported into $\Ne A \Gamma$ via reindexing with
$\tau : \Gamma \sublist \Delta$, in order to be applicable to the normal form
$\down[\Delta]A a$
reified from the semantic value $a$.

A direct extension of our presheaf semantics to positive types cannot
work.  For instance, with $\den \tyzero = \hat0$, simply
$\tfresh^{\tyzero}_{\cempty} : \bzero$
would give us an
inhabitant of the empty set, which means that reflection at the empty
type would not be definable.  Similarly, the setting
$\den{A + B} = \den A \hatplus \den B$ is refuted by
$\tfresh^{A + B}_{\cempty} : \den A (A{+}B) \bplus \den B (A{+}B)$
which would require us to make a decision of whether $A$ holds or $B$ holds
while only be given a hypothesis of type $A + B$.
Not even the usual interpretation of base types $\den o = \tNe\,o$
works in the presence of sums, as we would not be able to interpret
the term
$\tabs\,(\tcase\,\vv0\,\vv0\,\vv0) :
(o+o) \To o$ in our semantics, as $\Ne o (o+o)$ is empty.
What is needed are case distinctions on neutrals in the semantics,
allowing us the elimination of positive hypotheses before producing a
semantic value,
and we shall capture this capability in a strong monad $\C$
which can \emph{cover} the cases.

To recapitulate, a monad $\C$ on presheaves is first an endofunctor,
i.e., it maps any presheaf $\A$ to the presheaf $\C\,\A$ and any
presheaf morphism $f : \A \todot \B$ to the morphism
$\tmap[\C]f: \C\,\A \todot \C\;\B$ satisfying the functor laws for
identity and composition.  Then, there are natural transformations
$\treturn[\C] : \A \todot \C\,\A$ (unit) and
$\tjoin[\C] : \C\,(\C\,\A) \todot \C\,\A$ (multiplication)
satisfying
the monad laws.

We are looking for a cover monad $\C$ that
offers us these services:
\[
\begin{array}{lcl@{\quad}l}
\tabort^\C & : & \tNe\, \tyzero \todot \C\,\B
  & \mbox{case on absurd neutral} \\
\tcase^\C_\Gamma & : & \Ne {(A_1 + A_2)} \Gamma
  \to \C\,\B\,(\cext \Gamma {A_1})
  \to \C\,\B\,(\cext \Gamma {A_2})
  \to \C\,\B\,\Gamma
  & \mbox{case on neutral} \\
\trunNf[\C] & : & \C\,(\tNf A) \todot \tNf A
  & \mbox{run the monad ($\tNf$ only)} \\
\end{array}
\]
To make things concrete,
we shall immediately construct an instance of such a cover monad:
the free cover monad
$\Cov$ defined
as an inductive family with constructors
$\treturn[\Cov]$, $\tabort^{\Cov}$, and
$\tcase^{\Cov}$.  One can visualize an element $c : \Cov\,\A\,\Gamma$ as
binary case tree whose inner nodes ($\tcase$) are labeled by a neutral term
of sum type $A_1+A_2$ and its two branches by the context extensions $A_1$
and $A_2$, resp.  Leaves are either
labeled by a neutral term of empty type $\tyzero$ (see $\tabort$),
or by an element of $\A$ (see $\treturn$).  Functoriality amounts to
replacing the labels of the $\treturn$-leaves, and the monadic bind
(aka Kleisli extension) replaces
these leaves by further case trees.  (The uninspiring $\tjoin[\Cov]$
flattens a 2-level case tree, \ie, a case tree with case trees as
leaves, into a single one.)
Finally $\trunNf[\Cov]$ is a simple recursion on the tree, replacing $\tcase^{\Cov}$
and $\tabort^{\Cov}$
by the $\tcase$ and $\tabort$ constructions on normal forms, and
$\treturn[\Cov]$ by the identity.


Using the services of a generic cover monad $\C$, we can complete our semantics:
\[
\begin{array}{lcl}
  \den   \tybase & = & \C\,(\tNe\,\tybase) \\
  \up    \tybase & = & \treturn[\C] \\
  \down  \tybase & = & \trunNf[\C] \comp \tmap[\C]\tne \\
\end{array}
\qquad
\begin{array}{lcl}
  \den   \tyzero & = & \C\,\hat0 \\
  \up    \tyzero & = & \tabort^\C \\
  \down  \tyzero & = & \trunNf[\C] \comp \tmap[\C]\tmagic \\
\end{array}
\]
\[
\begin{array}{lcl}
  \den {A+B} & = & \C\,(\den A \hatplus \den B) \\
  \upG {A+B} \; u & = & \tcase^\C\,u
    \,(\treturn[\C]\,(\iota_1\,\tfresh^A_\Gamma))
    \,(\treturn[\C]\,(\iota_2\,\tfresh^B_\Gamma))
    \\
  \downG {A+B} & = & \trunNf[\C] \comp
    \tmap[\C] [\tinj_1 \comp \down A,\; \tinj_2 \comp \down B] \\
\end{array}
\]

All semantic types fulfill the weak sheaf condition aka weak pasting,
meaning there is
a natural transformation $\trun^A : \C\,\den A \todot \den A$ for any simple
type $A$.  In other words, we can \emph{run} the monad, pushing its
effects into $\den A$.  We proceed by induction on $A$.
Positive types $P$ are already monadic,
and $\trun^P$ is simply the $\tjoin$ of the monad $\C$.
At negative types we can recurse pointwise at a smaller type,
exploiting that values of negative types are essentially
(finite or infinite) tuples.
\[
\begin{array}{l@{~}l@{~~}c@{~~}l@{\qquad}l@{~~}c@{~~}l}
  \multicolumn 7 l {\trun^A ~:~ \C \den A \todot \den A} \\
  \trun^{\tyone} & c & = & \ttempty
    &
  \trun^{\tyzero} & = & \tjoin[\C]
    \\
  \trun^{A \times B} & c & = &
      (\trun^A\,(\tmap[\C]\;\pi_1\;c),\
       \trun^B\,(\tmap[\C]\;\pi_2\;c))
    &
  \trun^{A + B}   & = & \tjoin[\C]
  \\
  \trun^{A \To B} & c \; \tau \; a & = & \trun^B\,
      (\stmap[\C]\;(\lambda\;\tau'\,f.\;f\;\tid\;(\tren\;\tau'\;a))\;(\tren\;\tau\;c))
    &
  \trun^{o}      & = & \tjoin[\C]
\\
\end{array}
\]
For the case of function types $A \To B$, we require the monad $\C$ to be
\emph{strong}, which amounts to having
$\stmap[\C]_\Gamma\, \ell : \C\A\,\Gamma \to \C\,\B\,\Gamma$ already for
a ``local'' presheaf morphism $\ell : (\A \hatto \B)\Gamma$.
The typings are
$c : \C \den{A \To B} \Gamma$ and
$\tau : \Gamma \sublist \Delta$ and
$a : \den A \Delta$, and now we want to apply every function
$f : \den{A \To B}$ in the cover $c$ to argument $a$.  Clearly,
$\tmap[\C]$ is not applicable since it would expect a
global presheaf morphism
$\den{A \to B} \todot \den B$, \ie, something that works in \emph{any}
context.  However, applying to $a : \den A \Delta$ can only work in
context $\Delta$ or any extension $\tau' : \Delta \sublist \Phi$,
since we can transport $a$ to such a $\Phi$ via
$a' := \tren\; \tau'\; a : \den A \Phi$ but not to a context unrelated to
$\Delta$.  We obtain our input to $\trun^B$ of type $\C \den B \Gamma$
as an instance of $\stmap[\C]_\Gamma$ applied to the local presheaf morphism
$(\lambda \; \tau'\, f. \  f\;\tid\;a') :
\Delta \sublist \Phi \to  \den{A \To B}\Phi \to \den B \Phi$
and the transported cover $\tren\;\tau\;c : \C\den{A \To B}\Delta$.

We extend the type interpretation pointwise to contexts, \ie,
$\den \cempty = \hatone$ and
$\den{\cext \Gamma A} = \den \Gamma \hattimes \den A$
and obtain a natural projection function
$\tlookup\,(x : A \in \Gamma) : \den \Gamma \todot \den A$
from the semantic environments.
The evaluation function
\fbox{$\denty {t : A \from \Gamma}  : \den \Gamma \todot \den A$}
can now be defined by recursion on $t$.
Herein, the environment $\gamma$ lives in $\den \Gamma \Delta$, thus, $\dent t \gamma : \den A \Delta$.
\[
\begin{array}{lcl}
  \dent{\tunit}\gamma & = & \ttempty \\
  \dent{\tpair\,t_1\,t_2}\gamma & = &
    (\dent{t_1}\gamma,\ \dent{t_2}\gamma) \\
  \dent{\tabs\,t}\gamma & = & \fdent t \gamma \\
  \dent{\inj i t}\gamma & = & \iota_i \dent t \gamma \\
  \\
\end{array}
\quad
\begin{array}{lcl}
  \dent{\tvar\,x}\gamma & = & \tlookup\,x\,\gamma \\
  \dent{\prj i t}\gamma & = & \pi_i \dent t \gamma \\
  \dent{\tapp\,t\,u} \gamma & = &
     \dent t \gamma\;\tid\;\dent u \gamma \\
  \dent{\tcase\;u\;t_1\;t_2}\gamma & = &
     \dencase\;\dent u \gamma\;\fdent{t_1}\gamma\;\fdent{t_2}\gamma \\
  \dent{\tabort\,u}\gamma & = & \denabort\;\dent u \gamma \\
\end{array}
\]
For the interpretation of the binders $\tabs$ and $\tcase$ we use the
mutually defined $\fden{\_}$.
\[
\begin{array}{lcl}
  \fden{t : B \from \cext \Gamma A}
    & : & \den \Gamma \todot \den{A \To B} \\
    & = &
      \lambda\, (\gamma : \den \Gamma \Delta)
      \,(\tau : \Delta \sublist \Phi)
      \,(a : \den A \Phi)
      .\ \,
      \dent t {(\tren\,\tau\,\gamma,\ a)} \\
\end{array}
\]
The coproduct eliminations $\denabort$ and $\dencase$
targeting an arbitrary semantic type $\den B$ are definable thanks to
the weak sheaf property, \ie, the presence of pasting via $\trun^B$
for any type $B$, and strong functoriality of $\C$.
\[
\begin{array}{lcl}
  \denabort^B & : & \den \tyzero \todot \den B \\
  \denabort^B & = & \trun^B \comp \tmap[\C]\,\tmagic
\\[2ex]
  \dencase^B  & : & \den{A_1 + A_2}
    \todot \den{A_1 \To B}
    \todot \den{A_2 \To B}
    \todot \den{B} \\
  \dencase^B\;c\;f_1\;f_2 & = &
    \trun^B (\stmap[\C]\;(\lambda\,\tau.\ [f_1\,\tau,\ f_2\,\tau])\;c) \\
\end{array}
\]
To complete the normalization function
$\tnorm\,(t : A \from \Gamma) = \downG A \dent t {\idenv \Gamma}$
we define the identity environment $\idenv \Gamma : \den \Gamma \Gamma$,
which maps each free index to its corresponding unknown in the
semantics, by recursion on $\Gamma$:
\[
\begin{array}{lcl}
  \idenv \cempty & = & \ttempty \\
  \idenv {\cext \Gamma A} & = & (\tren\;\twk^A\,\idenv\Gamma, \  \tfresh^A_\Gamma)\\
\end{array}
\]

\subsection{Continuation monad}

As already observed by Filinski
\cite[Section~5.4]{filinski:semaccounttdpe}
\cite[Section~3.2]{filinski:tlca01},
normalization by evaluation can be carried out in the continuation
monad.
In our setting, we use a continuation monad $\CC$ on presheaves
defined as
\[
  \CC\;\J = \forall A.\ (\J \hatto \tNf\,A) \hatto \tNf\,A
  .
\]
The answer type of this continuation monad is always $\tNf$, however,
we are polymorphic in the simple type $A$ of normal forms we produce.%

Agda has been really helpful to produce the rather technical but
straightforward evidence that $\CC$ is a strong monad.
The method $\trunNf[\CC] : \CC\,(\tNf\,A) \todot \tNf\,A$
exists by definition, using the identity continuation $\tNf\,A \hatto \tNf\,A$.
In the following, we demonstrate that $\CC$
enables matching on neutrals:
\[
\begin{array}{l}
  \tabort^{\CC} \, (u : \Ne \tyzero \Gamma) ~:~  \CC\,\J\,\Gamma \\
  \tabort^{\CC}
    \; u
    \; (\tau : \Gamma \sublist \Delta)
    \; (k : (\J \hatto \tNf\,B)\Delta)
     ~=~
    \tabort\,(\tren\,\tau\,u)
\\[2ex]
  \tcase^{\CC}
     \; (u : \NeP {A_1 + A_2} \Gamma)
     \; (c_1 : \CC\,\J\,(\cext \Gamma {A_1}))
     \; (c_2 : \CC\,\J\,(\cext \Gamma {A_2}))
    ~:~ \CC\,\J\,\Gamma \\
  \tcase^{\CC}
    \; u
    \; c_1
    \; c_2
    \; (\tau : \Gamma \sublist \Delta)
    \; (k : (\J \hatto \tNf\,B)\Delta)
     ~=~
    \tcase\;(\tren\,\tau\,u)\;n_1\;n_2
    \bwhere \\
\qquad
  \begin{array}{l@{~~}c@{~~}l}
     n_i & : & \Nf B {(\cext \Delta {A_i})} \\
     n_i & = & c_i
      \; (\tlift^{A_i}\,\tau
            : \cext \Gamma {A_i} \sublist \cext \Delta {A_i})
      \; \left(\lambda
            \; (\tau' : \cext \Delta {A_i} \sublist \Phi)
            \; (j :  \J\,\Phi) .
            \ k
            \; (\twk^{A_i} \rcomp \tau')
            \; j
         \right)
  \end{array}
\end{array}
\]
The NbE algorithm using $\CC$ is comparable to Danvy's type-directed
partial evaluation \cite[Figure~8]{danvy:popl96}.  However, he uses
shift-reset style continuations which can be programmed in the
continuation monad, and relies on Scheme's \texttt{gensym} to produce
fresh variables names rather than using Kripke function space / presheaves.



\section{Normalization to Call-By-Push Value}
\label{sec:cbpv}

The placement of the monad $\C$
in the type semantics
of the previous section
is a bit wasteful:  Each positive type is
prefixed by $\C$.  In our grammar of normal forms, this corresponds to
the ability to perform case distinctions ($\tcase$, $\tabort$) at any
positive type $P$.  In fact, our type interpretation $\den A$
corresponds to the translation of call-by-name (CBN) lambda-calculus
into Moggi's monadic meta-language \cite{moggi:infcomp91,levy:hosc06}.

It would be sufficient 
to perform all necessary
case distinctions
when transitioning from a negative type to a positive type.
Introduction of the function type adds hypotheses to the context,
providing material for case distinctions, but introduction of positive
types does not add anything in that respect.  Thus, we could
\emph{focus} on positive introductions until we transition back to a
negative type.  Such focusing is present in the call-by-value (CBV)
lambda-calculus, where positive introductions only operate on values,
and variables stand only for values.  This structure is even more
clearly spelled out in Levy's call-by-push-value (CBPV)
\cite{levy:hosc06}, as it comes with a deep classification of types
into positive and negative ones.  In the following, we shall utilize
pure (\ie, effect-free) CBPV to achieve chaining of positive
introductions.

\subsection{Types and polarization}

CBPV calls positive types $P$ \emph{value types} $A$ and negative
types $N$
\emph{computation types} $\underline B$,
yet we shall stick to our terminology
which is common in publications on focalization.
However, we shall use $\tThunk$ for switch $\downarrow$ and $\tComp$ for
switch $\uparrow$.
\[
\begin{array}{lllrl@{\qquad}l}
  \PTy & \ni & P,Q & ::= & \patom
    \mid \tyone \mid P_1 \times P_2
    \mid \tyzero \mid P_1 + P_2
    \mid \Thunk N
  & \mbox{positive type}
\\
  \NTy & \ni & N,M & ::= & \natom
    \mid \top \mid N_1 \ntimes N_2
    \mid P \To N
    \mid \Comp P
  & \mbox{negative type}
\\
\end{array}
\]
CBPV uses $U$ for $\tThunk$ and $F$ for $\tComp$, however, we find
these names uninspiring unless you have good knowledge of the intended
model.  Further, CBPV employs labeled sums $\Sigma_I (P_i)_{i : I}$
and labeled records $\Pi_I (P_i)_{i : I}$ for up to countably infinite
label sets $I$ while we only have finite sums $(\tyzero,+)$ and
records $(\top,\ntimes)$.  However, this difference in not essential,
our treatment extends directly to the infinite case, since we are
working in type theory which allows infinitely branching inductive
types.   As a last difference, CBPV does not consider base types; in
anticipation of the next section, we add them as both positive atoms
($\patom$) and negative atoms ($\natom$).

Getting a bit ahead of ourselves, let us consider the
mutually defined interpretations
$\den P$ and $\den N$ of positive and negative types as presheaves.
\[
\begin{array}{lcl}
  \den{\tyone} & = & \hatone \\
  \den{P_1 \times P_2} & = & \den{P_1} \hattimes \den{P_2} \\
  \den{\tyzero} & = & \hatzero \\
  \den{P_1 + P_2} & = & \den{P_1} \hatplus \den{P_2} \\
  \den{\Thunk N} & = & \den N \\
  \den{\patom} & = & \patom{\in}\_ \\
\end{array}
\qquad
\begin{array}{lcl}
  \den{\top} & = & \hatone \\
  \den{N_1 \ntimes N_2} & = & \den{N_1} \hattimes \den{N_2} \\
  \\
  \den{P \To N} & = & \den P \hatto \den N \\
  \den{\Comp P} & = & \C \den P \\
  \den{\natom}  & = & \C (\tNe\, \natom) \\
\end{array}
\]
Semantically, we do not distinguish between positive and negative
products.  Notably, sum types can now be interpreted as plain
(pointwise) presheaf
sums.  The $\tThunk$ marker is ignored, yet $\tComp$, marking the
switch from the negative to the positive type interpretation, places
the cover monad.  Positive atoms, standing for value types without
constructors, are only inhabited by variables $x : \patom \in \Gamma$.
Negative atoms stand for computation types without own eliminations,
thus, their inhabitants stem only from eliminations of more complex
types, made from positive eliminations captured in $\C$ and negative
eliminations chained together as neutral $\tNe\,\natom$, which we
shall define below.
The method $\trunNf[\C]$ of cover monad $\C$ can be extended to
$\trun^N : \C \den N \to \den N$ for negative types $N$, by recursion
on $N$.  Informally speaking, this makes all negative types \emph{monadic}.

Contexts are lists of \emph{positive} types since in CBPV variables stand
for values.  Interpretation of contexts $\den \Gamma$ is again defined
pointwise $\den \cempty = \hatone$ and
$\den{\cext \Gamma P} = \den \Gamma \hattimes \den P$.

\subsection{Terms and evaluation}

Assuming a family $\Tm N \Gamma$ of terms of negative type $N$ in
context $\Gamma$, values \fbox{$v : \PValT P \Gamma$}
of positive type $P$ shall be constructed by the
following rules:
\begin{gather*}
  \nru{\tvar}{P \in \Gamma}{\PValT P \Gamma}
\qquad
  \nru{\tthunk}{\Tm N \Gamma}{\PValTP{\Thunk N}\Gamma}
\\[2ex]
  \nru{\punit}{}{\PValT{\tyone}\Gamma}
\qquad
  \nru{\ppair}
      {\PValT{P_1}\Gamma \qquad \PValT{P_2}\Gamma}
      {\PValTP{P_1 \times P_2}\Gamma}
\qquad
  \nru{\tinj_i}{\PValT{P_i}\Gamma}{\PValTP{P_1 + P_2}\Gamma}
\end{gather*}

The terms of pure CBPV are given by the inductive family \fbox{$\Tm N \Gamma$}.
It repeats the introductions and eliminations of negative types,
except that application is restricted to values.  Values of type
$\Thunk N$ are embedded via $\tforce$.
Further, values of type $P$ can embedded
via $\tret$, producing a term of type $\Comp P$.  Such terms are
eliminated by $\tbind$ which is, unlike the usual monadic bind, not
only available for $\tComp$-types but for arbitrary negative types
$N$.  This is justified by the monadic character of negative
types, by virtue of $\trun^N$.  Finally, there are eliminators
($\tsplit$, $\tcase$, $\tabort$) for values of positive product and
sum types.
\begin{gather*}
  \nru{\tret}
      {\PValTm P \Gamma}
      {\TmP {\Comp P} \Gamma}
\quad
  \nru{\tabs}
      {\Tm N {(\cext \Gamma P)}}
      {\TmP {P \To N} \Gamma}
\quad
  \nru{\npair}
      {\Tm {N_1} \Gamma \qquad \Tm {N_2} \Gamma}
      {\TmP {N_1 \ntimes N_2} \Gamma}
\quad
  \nru{\nunit}
      {}
      {\Tm \top \Gamma}
\\[2ex]
  \nru{\tforce}
      {\PValTmP {\Thunk N} \Gamma}
      {\Tm N \Gamma}
\qquad
  \nru{\tapp}
      {\TmP {P \To N} \Gamma \qquad \PValTm P \Gamma}
      {\Tm N \Gamma}
\qquad
  \nru{\tprj_i}
      {\TmP {N_1 \ntimes N_2} \Gamma}
      {\Tm {N_i} \Gamma}
\\[2ex]
  \nru{\tbind}
      {\TmP {\Comp P} \Gamma \qquad \Tm N {(\cext \Gamma P)}}
      {\Tm N \Gamma}
\qquad
  \nru{\tsplit}
    {\PValTmP {P_1 \times P_2} \Gamma
      \qquad \Tm N {(\cext {\cext \Gamma P_1} P_2)}
    }{\Tm N \Gamma}
\\[2ex]
  \nru{\tcase}
    {\PValTmP {P_1 + P_2} \Gamma
      \qquad \Tm N {(\cext \Gamma P_1)}
      \qquad \Tm N {(\cext \Gamma P_2)}
    }{\Tm N \Gamma}
\qquad
  \nru{\tabort}
    {\PValTm \tyzero \Gamma}
    {\Tm N \Gamma}
\end{gather*}

Interpretation of values
$\denty{ v : \PValTm P \Gamma } : \den \Gamma \todot \den P$
and terms
$\denty{ t : \Tm N \Gamma } : \den \Gamma \todot \den N$
is straightforward, thanks to the pioneering work of
Moggi \cite{moggi:infcomp91} and
Levy \cite{levy:hosc06}
put into the design of CBPV.
\[
\begin{array}{lcl}
  \\
  \dent{\tvar\,x}\gamma & = & \tlookup\,x\,\gamma \\
  \dent{\punit}\gamma   & = & \ttempty \\
  \dent{\ppair\,v_1\,v_2}\gamma & = &
    (\dent{v_1}\gamma,\ \dent{v_2}\gamma) \\
  \dent{\inj i v}\gamma & = & \iota_i \dent v \gamma \\
  \dent{\tthunk\,t}\gamma & = & \dent t \gamma \\
\end{array}
\quad
\begin{array}{lcl}
  \dent{\tabs\,t}\gamma
      & = & \fdent t \gamma \\
  \dent{\tapp\,t\,v} \gamma & = &
     \dent t \gamma\;\tid\;\dent v \gamma \\
  \dent{\nunit}\gamma & = & \ttempty \\
  \dent{\npair\,t_1\,t_2}\gamma & = &
    (\dent{t_1}\gamma,\ \dent{t_2}\gamma) \\
  \dent{\prj i t}\gamma & = & \pi_i \dent t \gamma \\
  \dent{\tforce\,v}\gamma & = & \dent v \gamma \\
\end{array}
\]
Since $\tThunk$ serves only as an embedding of negative into positive
types and has no semantic effect,
we interpret thunking and forcing by the identity.
The eliminations for positive types deal now only with values, thus,
need not reference the monad operations.
\[
\begin{array}{lcl}
  \dent{\tsplit\,v\,t}\gamma & = & (\lambda\,(a_1,a_2).\
    \dent{t}{(\gamma,a_1,a_2)} ) \; \dent v \gamma \\
  \dent{\tcase\,v\,t_1\,t_2}\gamma & = &
    [  \lambda a_1.\ \dent{t_1}{(\gamma,a_1)}
    ,\ \lambda a_2.\ \dent{t_2}{(\gamma,a_2)}
    ]\; \dent v \gamma \\
  \dent{\tabort\,v}\gamma & = & \tmagic\,\dent v \gamma \\
\end{array}
\]
The use of the monad is confined to $\tret$ and $\tbind$.
Note the availability of $\trun^\C : \C \den N \to \den N$
at any negative type $N$ for the interpretation of $\tbind$.
\[
\begin{array}{lcl}
  \dent{\tret\,v}\gamma & = & \treturn[\C] \; \dent v \gamma \\
  \dent{\tbind\,u\,t}\gamma & = & \trun^\C\,
    (\stmap[\C] \; \fdent t \gamma\; \dent u \gamma) \\
\end{array}
\]

\subsection{Normal forms and normalization}

Positive normal forms are values $v : \VNf P \Gamma$
referring only to atomic variables and
whose thunks only
contain negative normal forms.
\begin{gather*}
  \nru{\tvar}{\patom \in \Gamma}{\PValNf {\patom} \Gamma}
\qquad
  \nru{\tthunk}{\Nf N \Gamma}{\PValNfP{\Thunk N}\Gamma}
\\[2ex]
  \nru{\punit}{}{\PValNf{\tyone}\Gamma}
\qquad
  \nru{\ppair}
      {\PValNf{P_1}\Gamma \qquad \PValNf{P_2}\Gamma}
      {\PValNfP{P_1 \times P_2}\Gamma}
\qquad
  \nru{\tinj_i}{\PValNf{P_i}\Gamma}{\PValNfP{P_1 + P_2}\Gamma}
\end{gather*}

Neutral normal forms \fbox{$\Ne N \Gamma$} are negative eliminations
starting from a forced $\tThunk$ rather than from variables of negative
types (as those do not exist in CBPV).  However, due to normality
the $\tThunk$ cannot be a $\tthunk$, but only a variable
$\Thunk N \in \Gamma$.
\begin{gather*}
  \nru{\tforce}{\Thunk N \in \Gamma}{\Ne N \Gamma}
\qquad
  \nru{\tprj_i}
      {\NeP {N_1 \ntimes N_2} \Gamma}
      {\Ne {N_i} \Gamma}
\qquad
  \nru{\tapp}
      {\NeP {P \To N} \Gamma \qquad \PValNf P \Gamma}
      {\Ne N \Gamma}
\end{gather*}
Variables are originally introduced by either $\tabs$ or
the $\tbind$ing of a neutral of type $\Comp P$ to a new variable of type
$P$.  Variables of composite value type can be broken down by pattern
matching, introducing variables of smaller type.  These positive
eliminations plus $\tbind$ are organized in the
inductively defined strong monad \fbox{$\Cov$}.
\begin{gather*}
  \nru{\treturn}{\J\,\Gamma}{\Cover \J \Gamma}
\qquad
  \nru{\tbind}
      {\NeP {\Comp P} \Gamma \qquad \Cover \J (\cext \Gamma P)}
      {\Cover \J \Gamma}
\\[2ex]
  \nru{\tsplit}
      {P_1 \times P_2 \in \Gamma \qquad
       \Cover \J (\cext {\cext \Gamma {P_1}} {P_2})
      }
      {\Cover \J \Gamma}
\\[2ex]
  \nru{\tcase}
      {P_1 + P_2 \in \Gamma
       \qquad \Cover \J (\cext \Gamma {P_1})
       \qquad \Cover \J (\cext \Gamma {P_2})
      }
      {\Cover \J \Gamma}
\qquad
  \nru{\tabort}
      {\tyempty \in \Gamma}
      {\Cover \J \Gamma}
\end{gather*}
Finally, normal forms of negative types are defined as inductive
family \fbox{$\Nf N \Gamma$}.  They are generated by maximal negative
introduction ($\tabs$, $\npair$, $\nunit$) until a negative atom or
$\Comp P$ is reached.  Then, elimination of neutrals and variables is
possible through the $\Cov$ monad until an answer can be given in form
of a base neutral ($\tNe\,\natom$) or a normal value.
\begin{gather*}
  \nru{\tne}
      {\CoverP {\tNe\,\natom} \Gamma}
      {\Nf {\natom} \Gamma}
\qquad
  \nru{\tret}
      {\CoverP {\tVNf\,P}  \Gamma}
      {\NfP {\Comp P} \Gamma}
\\[2ex]
  \nru{\nunit}{}{\Nf \top \Gamma}
\qquad
  \nru{\npair}
      {\Nf {N_1} \Gamma \qquad \Nf {N_2} \Gamma}
      {\NfP {N_1 \ntimes N_2} \Gamma}
\qquad
  \nru{\tabs}
      {\Nf N {\cext \Gamma P}}
      {\NfP {P \To N} \Gamma}
\end{gather*}
We again can run the cover monad on normal forms, \ie, have
$\trunNf : \Cov\,(\tNf N) \todot \tNf N$, which extends to
negative semantic values
$\trun : \Cov \den N \todot \den N$.

Reification $\down P : \den P \todot \tVNf\,P$ at positive types $P$ produces
a normal value, and $\down N : \den N \todot \tNf\,N$ at negative
types $N$ a
normal term.  During reification of function types $P \To N$ in
context $\Gamma$ we need to embed a fresh variable
$x : P \in (\cext \Gamma P)$ into $\den P$, breaking down $P$ to
positive atoms $\patom$ and negative remainders $\Thunk N$.
However, in $\den P$ we do not have case analysis available, thus,
positive reflection $\upG P : P \in \Gamma \to \Cover {\den P} \Gamma$
needs to run in the monad.  Luckily, as $\down N$ produces a normal
form, monadic intermediate computations are permitted under a final
$\trunNf$.  Negative reflection $\up N : \tNe\,N \todot \den N$ is as
before generalized from variables to neutrals, to handle the breaking
down of $N$ via eliminations.
In the following definition of reflection we use the abbreviation
\fbox{\(
  \tfresh^P_\Gamma
    = \up[\cext \Gamma P] P \vv0
    : \Cover{\den P}{(\cext \Gamma P)}
\)}.
\[
\def\arraystretch{1.3}
\begin{array}{l@{\,}l@{~~}c@{~~}l}
  \multicolumn 2 l {\upG P} & : & P \in \Gamma \to \Cov\,\den P \, \Gamma\\
  \upG {\patom}         & x & = & x \\
  \upG {\tyone}         & x & = & \treturn\;\ttempty \\
  \upG {P_1 \times P_2} & x & = & \tsplit\;x\,
    \left(
     (\up[\cext{\cext \Gamma {P_1}}{P_2}]{P_1}\,\vv1)
     \star
     (\up[\cext{\cext \Gamma {P_1}}{P_2}]{P_2}\,\vv0)
    \right) \\
  \upG{\tyzero}         & x & = & \tabort\;x \\
  \upG{P_1 + P_2}       & x & = & \tcase\;x
    \; (\tmap\;\iota_1\;\tfresh^{P_1}_\Gamma)
    \; (\tmap\;\iota_2\;\tfresh^{P_2}_\Gamma)
    \\
  \upG{\Thunk N} & x & = & \treturn\,(\upG N (\tforce\,x)) \\
\end{array}
\,
\begin{array}{lcl}
  \down P & : & \den P \todot \tVNf\,P \\
  \down {\patom} & = & \tvar \\
  \downG {\tyone} () & = & \punit \\
  \downG {P_1 \times P_2}(a_1,a_2) & = & \ppair
    \,(\downG {P_1} a_1)
    \,(\downG {P_2} a_2)
    \\
  \down {\tyzero} & = & \tmagic \\
  \down {P_1 + P_2} & = &
    [  \tinj_1 \comp \down{P_1}
    ,\ \tinj_2 \comp \down{P_2}
    ] \\
  \down {\Thunk N} & = & \tthunk \circ \down N \\
\end{array}
\]
Reflection at positive pairs uses monoidal functoriality
$\C\,\A_1 \todot \C\,\A_2 \todot \C\,(\A_1 \hattimes \A_2)$
called $\star$ by McBride and Paterson \cite[Section 7]{mcbridePaterson:applicative}
which for monads $\C$ can be defined by
$c_1 \star c_2 =
   \tjoin \; (\tmap\;(\lambda\,a_1.\
   \tmap\;(\lambda\,a_2.\ (a_1,a_2))
    \;c_2)
    \;c_1)$.

For negative types, reflection and reification works as before:
\[
\def\arraystretch{1.3}
\begin{array}{l@{\,}l@{~~}c@{~~}l}
  \multicolumn 2 l {\up N} & : & \tNe\,N \todot \den N \\
  \upG {\Comp P} & u & = & \tbind\,u\,\tfresh^P_\Gamma \\
  \upG \natom    & u & = & \treturn\,u \\
  \upG \top      & u & = & \ttempty \\
  \upG {N_1 \ntimes N_2} & u & = & \left(
    \upG {N_1} \; (\prj 1 u)
    ,\
    \upG {N_2} \; (\prj 2 u)
    \right) \\
\end{array}
\
\begin{array}{l@{\,}l@{~~}c@{~~}l}
  \multicolumn 2 l {\down N} & : & \den N \todot \tNf\;N \\
  \downG {\Comp P} & c  & = & \tmap\;(\down P)\;c \\
  \downG \natom    & c  & = & \tne\,c \\
  \downG \top      & () & = & \nunit \\
  \downG {N_1 \ntimes N_2} & (b_1,b_2) & = & \npair
    \, (\downG {N_1} \; b_1)
    \, (\downG {N_2} \; b_2)
    \\
\end{array}
\]
Reflection for function types is also unchanged, except that $\tapp$
expects a value argument now.
\[
\def\arraystretch{1.3}
\begin{array}{l@{\,}l@{~~}c@{~~}l}
  \upG {P \To N} & u & = & \lambda
    \;(\tau : \Gamma \sublist \Delta)
    \;(a : \den P \Delta)
    . \
    \up[\Delta] N \left( \tapp\,(\tren\,\tau\,u)\,(\down[\Delta] P a) \right)
    \\
  \downG {P \To N} & f & = & \tabs \left( \trunNf \left(
    \stmap
      \left(
        \lambda
          \;(\tau : (\cext \Gamma P) \sublist \Delta)
          \;(a : \den P \Delta)
          . \
          f\; (\twk^P \rcomp \tau)\; a
        \right)
      \,\tfresh^P_\Gamma
    \right) \right)
    \\
\end{array}
\]
The identity environment $\idenv \Gamma : \Cover {\den \Gamma} \Gamma$
can only be generated in the monad, due to monadic positive
reflection.
\[
\begin{array}{lcl}
  \idenv{\cempty} & = & \treturn\,() \\
  \idenv{\cext \Gamma P} & = &
    (\tren\,\twk^P\,\idenv\Gamma) \star \tfresh^P_\Gamma \\
\end{array}
\]
Putting things together, we obtain the normalization function
\[
  \tnorm\,(t : \Tm N \Gamma) = \trunNf\,
    \left(
      \tmap\,\left(\down N \comp \denty t\right)\,\idenv\Gamma
    \right)
.
\]

Taking stock, we have arrived at normal forms that eagerly introduce
($\tNf$) and eliminate ($\tNe$) negative types and also eagerly
introduce positive types ($\tVnf$).  However, the elimination of
positive types is still rather non-deterministic.
It is possible to only partially break up a composite positive type
and leave smaller, but still composite positive types for later
pattern matching.
The last refinement, chaining also the positive eliminations, will be
discussed in the following section.

\section{Focused Intuitionistic Propositional Logic}
\label{sec:fipl}
\label{sec:pol}



Polarized lambda-calculus
\cite{zeilberger:PhD,espiritoSanto:entcs17}
is a focused calculus, it eagerly employs
so-called \emph{invertible} rules: the introduction rules for negative types and the
elimination rules for positive types.  As a consequence of the latter,
variables are either of atomic or negative type $H ::= \patom \mid N$.
Contexts $\Gamma,\Delta$ are lists of $H$s.

To add a variable of positive type $P$ to the context, we need to
break it apart until only atoms and negative bits remain.  This is
performed by maximal pattern matching, called the left-invertible
phase of focalization.%
\footnote{Filinski \cite[Section~4]{filinski:tlca01} achieves maximal pattern
matching through an additional, ordered context $\Theta$ for positive
variables which are eagerly split.}
We express maximal pattern matching on $P$ as a strong functor
$\Add P$ in the category of presheaves,
mapping a presheaf $\J$ (``judgement'') to $\Add P \J$
and a presheaf morphism $f : (\J \hatto \K)\Gamma$ to
$\stmap[\Add P]_\Gamma f : \AddHyp P \J \Gamma \to \AddHyp P \K \Gamma$.
For arbitrary $\J$ and $\Gamma$, the family \fbox{$\AddHyp P \J \Gamma$} is
inductively constructed by the following rules:
\begin{gather*}
  \nru{\phyp}
      {\J\,(\cext \Gamma \patom)}
      {\AddHyp {\patom} \J \Gamma}
\qquad
  \nru{\nbranch}
      {}
      {\AddHyp {\tyzero} \J \Gamma}
\qquad
  \nru{\bbranch}
      {\AddHyp {P_1} \J \Gamma \qquad
       \AddHyp {P_2} \J \Gamma }
      {\AddHyp {P_1 + P_2} \J \Gamma}
\\[2ex]
  \nru{\nhyp}
      {\J\,(\cext \Gamma N)}
      {\AddHyp {\Thunk N} \J \Gamma}
\qquad
  \nru{\nsplit}
      {\J\,\Gamma}
      {\AddHyp {\tyone} \J \Gamma}
\qquad
  \nru{\bsplit}
      {\AddHyp {P_1} {\left(\Add{P_2}\,\J\right)} \Gamma}
      {\AddHyp {P_1 \times P_2} \J \Gamma}
\end{gather*}
Note the recursive occurrence of $\Add{P_2}$ as argument to
$\Add{P_1}$ in $\bsplit$, which makes $\Add P$ a nested datatype
\cite{bird:nested}.  Agda supports such nested inductive types; but
note that $\Add{P}$ is uncontroversial, since it could also be defined
by recursion on $P$.   It is tempting to name $\bsplit$
``$\tjoin$'' and $\nsplit$ ``$\treturn$''
since $\Add P$ is a graded monad on the monoid $(1,\times)$
of product types;
however, this coincidence shall not matter for our further considerations.

Focalization is a technique to remove don't-care non-determinism from
proof search, and as such, polarized lambda calculus is foremost a
calculus of normal forms.
These normal forms are given by four mutually defined inductive
families of presheaves $\tVnf\,P$,
$\tNe\,N$, $\Cov\,\J$,
and $\tNf\,N$.  As they are very similar to the CBPV normal forms
given in the last section, we only report the differences.
Values \fbox{$v : \PVal P \Gamma$} are unchanged, they can refer to
atomic positive hypotheses ($\pvar$) and normal $\tthunk$s.
Neutrals \fbox{$\Ne N \Gamma$} start with a negative variable instead
of with $\tforce$, as forcing thunks is already performed in $\nhyp$
when adding hypotheses of $\tThunk$ type.
%
The normal forms \fbox{$\Nf N \Gamma$} of negative type are unchanged
with the exception that pattern matching happens eagerly in $\tabs$,
by virtue of $\Add P$.
\[
  \nru{\nvar}{N \in \Gamma}{\Ne N \Gamma}
\qquad
\qquad
  \nru{\tabs}
      {\AddHypP P {\tNf\, N} \Gamma}
      {\NfP {P \To N} \Gamma}
\]
The Cover monad \fbox{$\Cover \J \Gamma$}
lacks constructors $\tsplit$, $\tcase$ and $\tabort$
since the pattern matching is taken care of by $\Add P$.
\begin{gather*}
  \nru{\treturn}{\J\,\Gamma}{\Cover \J \Gamma}
\qquad
  \nru{\tbind}
      {\NeP {\Comp P} \Gamma \qquad \AddHypP P {\Cov\,\J} \Gamma}
      {\Cover \J \Gamma}
\end{gather*}

All these inductive families are presheaves, due to factored
presentation using $\Add P$ and $\Cov$ the proof is not a simple
mutual induction.  Yet, in Agda, the generic proof goes through using a sized
typing for these inductive families.  Similarly, defining the
$\tjoin$ for monad $\Cov$ relies on sized typing \cite{abel:PhD}.
\[
\begin{array}{lll}
  \tjoin & : & \forall i.\
    \Cov[i]\,(\Cov[\infty]\J) \todot \Cov[\infty]\J  \\
  \tjoin[i+1]\,(\treturn[i]\;c) & = & c \\
  \tjoin[i+1]\,(\tbind^i\;t\;k)
    & = & \tbind^\infty\;
     (t : \Ne P \Gamma)\;
     \left(\stmap[\Add P]_\Gamma\;\tjoin[i]\;
       \left(k : \AddHypP P {\Cov[i]\J} \Gamma
       \right)
     \right)
\end{array}
\]
Herein, we used the sized typing of the constructors of $\Cov$:
\[
\begin{array}{lll}
  \treturn & : & \forall i.\ \J \todot \Cov[i+1] \J \\
  \tbind  & : & \forall i.\
    \tNe\,(\Comp P) \todot \Add P\,(\Cov[i]\J) \todot \Cov[i+1]\J \\
\end{array}
\]

Due to the eager splitting of positive hypotheses, reflection at type $P$
now lives in the graded monad $\Add P$ rather than $\Cov$.  Further,
as pattern matching may produce $n \geq 0$ cases, reflection cannot simply
produce a single positive semantic value; instead, one such value is needed
for every branch.  We implement
\fbox{$\reflect P : (\den P \hatto \J) \todot \Add P \J$}
as a higher-order function expecting a continuation $k$ which is
invoked for each generated branch with the semantic value of type $P$
constructed for this branch.
\[
\begin{array}{l@{~}lcl}
  \reflect {\patom} & k & = &
    \phyp \left( k\;\twk^{\patom}\,(\pvar\,\tzero) \right) \\
  \reflect {\Thunk N} & k & = &
    \nhyp \left( k\;\twk^N \left( \up N (\nvar\,\tzero) \right) \right)\\
  \reflect {\tyzero} & k & = & \nbranch \\
  \reflect {P_1 + P_2} & k & = & \bbranch
    \,\left( \treflect^{P_1}\,(\lambda\,\tau. \ k\,\tau \comp \iota_1) \right)
    \,\left( \treflect^{P_2}\,(\lambda\,\tau. \ k\,\tau \comp \iota_2) \right)
    \\
  \reflect{\tyone} & k & = & \nsplit\;(k\;\tid\;\ttempty) \\
  \reflect{P_1 \times P_2} & k & = & \bsplit
    \,\left(
      \treflect^{P_1} \left( \lambda\,\tau_1\,a_1. \
      \treflect^{P_2} \left( \lambda\,\tau_2\,a_2. \
        k\, (\tau_1 \rcomp \tau_2)\, (\tren\,\tau_2\,a_1, \ a_2)
        \right)
      \right)
    \right)
\end{array}
\]
Reflecting at a positive atomic type $\patom$ is the regular ending of a
reflection pass: we call continuation $k$ with a fresh variable
$\pvar\,\tzero$ of type $\patom$, making space for the variable
using $\twk^{\patom}$.
In case we end at type $\Thunk N$, we add a new variable
$\nvar\,\tzero$ of type $N$ and pass it to $k$, after full
$\eta$-expansion via $\up N$.  Two more endings are possible:
At type
$\tyzero$, we have reached an absurd case, meaning that no
continuation is necessary since we can conclude with
\emph{ex falsum quod libet}.
At type $\tyone$, there is no need to add a new variable, as values
of type $\tyone$ contain no information.  We simply pass the unit
value $\ttempty$ to $k$ in this case.
Reflecting at $P_1 + P_2$ generates two branches, which may result in
several uses of the continuation $k$.
In the first branch, we recursively reflect at $P_1$.  Its
continuation will receive a semantic value in $\den{P_1}$ which we
inject via $\iota_1$ into $\den{P_1 + P_2}$ to pass it to $k$.  The
second branch proceeds analogously.
Finally reflecting at $P_1 \times P_2$ means we first have to analyze
$P_1$, and in each of the generated branches we continue to analyze
$P_2$.  Thus $\reflect{P_2}$ is passed as a continuation to
$\reflect{P_1}$.  Each reflection phase gives us a semantic value
$a_i$ of type $P_i$, which we combine to a tuple before passing it to
$k$.  Note also that the context extension $\tau_1$ created in the
first phase needs to be composed with the context extension $\tau_2$
of the second phase to transport $k$ into the final context.  Further,
the value $a_1$ was constructed relative to the target of
$\tau_1$ and still needs
to be transported with $\tau_2$ before paired up with $a_2$.

The method $\reflect P$ replaces previous uses of $\tfresh^P$ in
reflection and reification at negative types.
\[
\begin{array}{l@{~}lcl}
  \downG {P \To N} & f & = & \tabs
    \left(
      \reflect[\Gamma] P
        \left(
          \lambda \, (\tau : \Gamma \sublist \Delta) \; a .\
          \down[\Delta] P (f \; \tau \; a)
        \right)
    \right)
  \\
  \upG {\Comp P} & u & = & \tbind \; u
    \left(
      \reflect[\Gamma] P
        \left(
          \lambda \; \tau \; a .\
          \treturn\; a
        \right)
    \right)
\end{array}
\]
Due to the absence of composite positive types in contexts,
the identity environment $\idenv \Gamma$ can be built
straightforwardly using negative reflection.
\[
\begin{array}{lcl}
  \idenv{\cempty} & = & \ttempty \\
  \idenv{\cext \Gamma {\patom}} & = &
    (
      \tren\;\twk^{\patom}\;\idenv\Gamma
      , \
      \pvar\,\tzero
    ) \\
  \idenv{\cext \Gamma N} & = &
    (
      \tren\;\twk^N\;\idenv\Gamma
      , \
      \upG N (\nvar\,\tzero)
    ) \\
\end{array}
\]

The terms \fbox{$\Tm N \Gamma$} of the focused lambda calculus are the
ones of CBPV minus the positive eliminations ($\tsplit$, $\tcase$,
$\tabort$), the added negative variable rule ($\nvar$) and the
necessary changes to the binders $\tabs$ and $\tbind$.
\begin{gather*}
  \nru{\nvar}{N \in \Gamma}{\Tm N \Gamma}
\qquad
  \nru{\tabs}
      {\AddHypP P {\tTm\, N} \Gamma}
      {\TmP {P \To N} \Gamma}
\qquad
  \nru{\tbind}
      {\TmP {\Comp P} \Gamma \qquad \AddHypP P {\tTm\,N} \Gamma}
      {\Tm N \Gamma}
\end{gather*}
Term interpretation
\fbox{$\denty\_ : \Tm N \Gamma \to \den \Gamma \todot \den N$}
shall be as for CBPV except that we need to exchange the
interpretation function for binders \fbox{$\fden\_ : \Tm N {(\cext
    \Gamma P)} \to \den \Gamma \todot \den{P \To N}$}.  Since a binder
for $P$ performs a maximal splitting on $P$ and takes the form of a
function defined by a case (and split) tree,
applying it to a value $v$ of type $P$ amounts to a complete
matching of $v$ against the case tree, and bind the remaining
atomic and negative crumbs.  This matching can be defined for a
generic evaluation function of type
$\Evvv \J \Delta \Gamma = \den \Gamma \Delta \to \J\,\Delta$.
\[
\begin{array}{l@{~}l@{~}l@{~}lcl}
  \multicolumn 6 {l} {\tmatch ~~:~~ \den P \Delta \to
    \AddHyp P {\Evv \J \Delta} \todot \Evv \J \Delta }\\
  \tmatch & x & (\phyp\,e) & \gamma & = & e\,(\gamma,\; \treturn\,x) \\
  \tmatch & b & (\nhyp\,e) & \gamma & = & e\,(\gamma,\; b) \\
  \tmatch & \ttempty & (\nsplit\,e) & \gamma & = & e\,\gamma \\
  \tmatch & (a_1,a_2) & (\bsplit\,e) & \gamma & = &
    \tmatch\;a_1\;(\tmap[\Add P]\,(\tmatch\;a_2)\;e)\;\gamma \\
  \tmatch & a & \nbranch & \gamma & = & \tmagic\;a \\
  \tmatch & (\iota_1\;a_1) & (\bbranch\;e_1\;e_2) & \gamma & = &
    \tmatch\;a_1\;e_1\;\gamma \\
  \tmatch & (\iota_2\;a_2) & (\bbranch\;e_1\;e_2) & \gamma & = &
    \tmatch\;a_2\;e_2\;\gamma \\
\end{array}
\]
With instantiations $\J = \den N$ and
$\denty\_ : \Tm N \todot \Evv {\den N} \Delta$,
the interpretation $\fden t$
of binder $t : \AddHypP P {\tTm\,N} \Gamma$
is defined as follows:
\[
  \fdent{t}
    {(\gamma : \den \Gamma \Delta)}
    \, (\tau : \Delta \sublist \Phi)
    \, (a : \den P \Phi)
    ~~=~~
    \tmatch
      \; a
      \; (\tmap[\Add P]\;\denty\_\;t)
      \; (\tren\,\tau\,\gamma)
\]
This completes the definition of the normalization function
$\tnorm\, (t : \Tm N \Gamma)
  = \down N \dent t {\idenv\Gamma}$.

\section{Conclusion}
\label{sec:concl}

We have implemented NbE for CBPV and polarized lambda calculus
formulated with intrinsically well-typed syntax and presheaf
semantics.  As a side result, we have proven semantically
that the normal forms of both systems are logically complete, \ie,
each derivable judgement $\Gamma \der N$ has a normal derivation.
It remains to show that NbE for these calculi is also computationally
sound and complete, i.e., the computational behavior of term and
normal form should agree, and normalization should decide a suitable
equational theory on terms.

Additionally, a natural question to investigate is whether known CBN
and CBV NbE algorithms can be obtained from our NbE algorithms by
embedding simply-typed lambda calculus into our polarized calculi,
using known CBN and CBV translations.  Further, we would like to study
the NbE algorithm for STLC arising from the optimal translation, i.e.,
the one inserting a minimal amount of $\tThunk$ and $\tComp$ transitions.

\bibliography{auto-fscd19}

\appendix

\section{Equational Theory for STLC with Weak Sums}
\label{sec:eq}

Term equality\fbox{$t \cong t'$} for $t,t' : A \from \Gamma$ is the
least congruence over the following axioms, which we have grouped into
$\beta$ (computation), $\eta$ (extensionality), and $\pi$ (permutation)
rules.  We may refer to subgroups, like $\eta^N$ for the negative
$\eta$-rules ($\retato$, $\retatimes$, $\retaone$).  Note that all
typing restrictions are implicit in the well-typedness assumption,
\eg, $t \cong \tunit$ presupposes $t : \tyone \from \Gamma$
since $\tunit : \tyone \from \Gamma$.
\begin{gather*}
  \nru{\rbetato}{}{\tapp\,(\tabs\,t)\,u \cong \subone t u}
\qquad
  \nru{\rbetatimes}{}{\prjP i {\tpair\,t_1\,t_2} \cong t_i}
\qquad
  \nru{\rbetaplus}{}{\tcase\,(\inj i t)\,t_1\,t_2 \cong t_i}
\\[2ex]
  \nru{\retato}{}{t \cong
    \tabs\,(\tapp\,(\tren\,\twk\,t)\,\vv0)}
\quad
  \nru{\retatimes}{}{t \cong \tpair\,(\prj 1 t)\,(\prj 2 t)}
\quad
  \nru{\retaone}{}{t \cong \tunit}
\\[1ex]
  \nru{\retaplus}{}{t \cong
    \tcase\,t
      \,(\inj 1 {\vv0})
      \,(\inj 2 {\vv0})
    }
\qquad
  \nru{\retazero}{}{t \cong \tabort\,t}
\\[2ex]
  \nru{\rpi{\To}{\tyempty}}
    {}{\tapp\,(\tabort\,t)\,u \cong \tabort\,t}
\qquad
  \nru{\rpi{\To}{+}}
      {t_i' = \tapp\,t_i\,(\tren\,\twk\,u)}
      {\tapp\,(\tcase\;t\;t_1\;t_2)\,u \cong \tcase\;t\;t_1'\;t_2'}
\\[1ex]
  \nru{\rpi{\times}{\tyempty}}
    {}{\prjP i {\tabort\,t} \cong \tabort\,t}
\qquad
  \nru{\rpi{\times}{+}}
    {}{\prjP i {\tcase\;t\;t_1\;t_2} \cong
         \tcase\,t\,(\prj i {t_1})\,(\prj i {t_2})}
\\[2ex]
  \nru{\rpi{+}{\tyempty}}
    {}{\tcase\,(\tabort\,t)\,t_1\,t_2 \cong \tabort\,t}
\qquad
  \nru{\rpi{+}{+}}
      {t_i' = \tcase\;t_i\;u_1'\;u_2'
       \quad
       u_j' = \tren\,(\tlift\,\twk)\,u_j
      }
      {\tcase\,(\tcase\;t\;t_1\;t_2)\,u_1\,u_2 \cong
        \tcase\;t\;t_1'\;t_2'}
    \
\\[1ex]
  \nru{\rpi{\tyempty}{\tyempty}}
    {}{\tabort\,(\tabort\,t) \cong \tabort\,t}
\qquad
  \nru{\rpi{\tyempty}{+}}
    {}{\tabort\,(\tcase\;t\;t_1\;t_2) \cong
      \tcase\,t\,(\tabort\,t_1)\,(\tabort\,t_2)}
\end{gather*}
The different law classes contribute to the shape and completeness of
the normal forms in the following way:
Thanks to the $\eta$-laws, a normal form of negative type is always an
introduction.  A normal form of sum type ($+$, $\tyzero$) is always a
case tree whose leaves are injections.  A normal form of atomic type
($o$) is always case tree whose leaves are neutral.
The $\beta$ and $\pi$ laws determine the shape of neutrals.
Thanks to the $\beta$-laws,
the principal argument of normal eliminations cannot be an
introduction.  Because of the negative permutation rules
$\rpi N P$ it cannot be positive elimination
($\tcase$, $\tabort$) either, which leaves only negative eliminations
in neutrals ($\tapp$, $\tprj$).  Finally the positive permutation
rules $\rpi P P$ guarantee that having neutral scrutinees in case
trees is sufficient.

\end{document}